\documentclass[usenames,dvipsnames]{scrartcl}
\usepackage{ stmaryrd }
\usepackage{ bm }
\usepackage[utf8]{inputenc}
\usepackage{bbm}
\usepackage[colorinlistoftodos]{todonotes}
\usepackage{algorithm}
\usepackage[noend]{algpseudocode}
\usepackage{mathtools}
\usepackage[toc,page]{appendix}
\usepackage{amsmath,amsfonts,amssymb,amsthm,mathrsfs} \usepackage[a4paper,vmargin={2.4cm,2.8cm},hmargin={2.2cm,2.2cm}]{geometry} 
\usepackage[labelfont={bf}, margin=1cm]{caption} 
\usepackage{graphicx,graphics} 
\usepackage{epsfig} 
\usepackage{latexsym} 
\usepackage[english]{babel} 
\usepackage{enumerate}
\usepackage{changepage}
\usepackage{ytableau} 
\usepackage[numbers]{natbib}
\ytableausetup{smalltableaux=true}

\DeclareMathOperator{\Tr}{Tr}
\DeclareMathOperator{\Diag}{Diag}
\DeclareMathOperator{\Vand}{Vand}
\DeclareMathOperator{\Eigen}{Eigen}

\usepackage{thmtools}

\usepackage{geometry}

\usepackage{authblk}
\title{Asymptotic behavior of the multiplicative counterpart of the Harish-Chandra integral and the $S$-transform}
\author[,1,2]{Pierre Mergny\thanks{\texttt{mergny.pierre@gmail.com}}}
\author[3]{Marc Potters}
\affil[1]{Chair of Econophysics $\&$ Complex Systems, Ecole polytechnique, 91128 Palaiseau Cedex, France}
\affil[2]{LPTMS,  CNRS,  Univ.   Paris-Sud,  Université  Paris-Saclay,  91405  Orsay,  France}
\affil[3]{Capital Fund Management, 23 rue de l’Université, 75007 Paris, France}
\date{}
\begin{document}
\maketitle

\begin{abstract}
In this note, we study the asymptotic of spherical integrals, which are analytical extension in index of the normalized Schur polynomials for $\beta =2$ , and of Jack symmetric polynomials otherwise. Such integrals are the multiplicative counterparts of the Harish-Chandra-Itzykson-Zuber (HCIZ) integrals, whose asymptotic are given by the so-called $R$-transform when one of the matrix is of rank one. We argue by a saddle-point analysis that a similar result holds for all $\beta >0$  in the multiplicative case, where the asymptotic is governed by the logarithm of the $S$-transform.  As a consequence of this result one can calculate the asymptotic behavior of  complete homogeneous symmetric polynomials. 
\end{abstract}

\section{Introduction}

\subsection{Free probability transforms}

Let $\mathbf{A}_N$ (respectively $\mathbf{B}_N$)  be a  random orthogonal/unitary/symplectic invariant matrices such that their empirical spectral distributions $\mu_{\mathbf{A}_N}$ (resp.  $\mu_{\mathbf{B}_N}$)  converge in the large $N$ limit towards a deterministic measure $\mu_A$ (resp. $\mu_B$)\footnote{we also assume that the minimum and maximum of the empirical spectral measure converge towards the edge of the limiting distribution}, then it is well known from free probability that the empirical distribution of the free sum $\mathbf{C}_N = \mathbf{A}_N + \mathbf{G} \mathbf{B}_N \mathbf{G^*}$, with $\mathbf{G}$ taken uniformly distributed  in the $N$-orthogonal/unitary/symplectic group, and $^*$ denotes the conjugation operation, converges towards the \emph{free additive convolution}  of $\mu_A$ and $\mu_B$, denoted by $\mu_A \boxplus \mu_B$. The $R$-transform defined by: 

\begin{align}
\mathcal{R}_{\mu}(z) := \mathcal{G}_{\mu}^{(-1)}\left( z\right) - \frac{1}{z}
\end{align}
where $\mathcal{G}_{\mu}(z) := \int \frac{1}{z-x} \mu(dx)$  is the \emph{Stieltjes transform} and $f^{(-1)}(.)$ denotes the functional inverse of the function $f(.)$, linearizes this convolution: 

\begin{equation}
\mathcal{R}_{\mu_A \boxplus \mu_B}(z) = \mathcal{R}_{\mu_A}(z) + \mathcal{R}_{\mu_B}(z)
\end{equation}
Similarly, for two positive definite orthogonal/unitary/symplectic invariant matrices $\mathbf{A}_N$ and $\mathbf{B}_N$, the limiting spectral distribution of the free product $\mathbf{C}_N = \sqrt{\mathbf{A}_N} \mathbf{G} \mathbf{B}_N \mathbf{G}^* \sqrt{\mathbf{A}_N}$ is given by \emph{the free multiplicative convolution} $\mu_A \boxtimes \mu_B$ and we have: 

\begin{align}
\tilde{\mathcal{S}}_{\mu_A \boxtimes \mu_B}(z) = \tilde{\mathcal{S}}_{\mu_A}(z) \,  \tilde{\mathcal{S}}_{\mu_B}(z)
\end{align}
where $\tilde{\mathcal{S}}(.)$ is the (modified) $\mathcal{S}$-transform\footnote{the usual $\mathcal{S}$-transform is defined as $ \frac{1}{\tilde{\mathcal{S}}}$ }  defined by: 
\begin{align}
\label{ModS-tsfrm}
\tilde{\mathcal{S}}_{\mu}(z) = \frac{z}{z+1} \mathcal{T}_{\mu}^{(-1)}(z)
\end{align}
with $\mathcal{T}_{\mu}(z):= z\mathcal{G}_{\mu}(z) -1$.  In particular its logarithm plays the role of the $R$-transform since it linearizes the free multiplicative convolution. 

\subsection{Harish-Chandra-Itzykson-Zuber Integrals and free sum}

The \emph{Harish-Chandra-Itzykson-Zuber} (HCIZ in the following)  integral \cite{HarishChandra1957} \cite{Itzykson1980} \cite{McSwiggen2018} is defined in the RMT setting as: 

\begin{align}\label{HCIZdef}\mathcal{I}^{(\beta)} \left( \mathbf{A}_N , \mathbf{Z}_N \right): = \int \mathcal{D}\mathbf{G} \, e^{\Tr \mathbf{A}_N \mathbf{G} \mathbf{Z}_N \mathbf{G}^*} && \left(\text{for } \beta = 1,2,4\right) 
\end{align}

where the integral is over the orthogonal (respectively unitary, symplectic) group for $\beta =1$ (respectively $\beta =2$ , $\beta=4$) and $\mathcal{D}\mathbf{G}$ is the corresponding Haar measure normalized to unity $\int \mathcal{D}\mathbf{G} = 1$. Such integral has been proven to be of high interest in lattice gauge theory \cite{Itzykson1980}, algebraic geometry \cite{Goulden2014} \cite{novak2020complex} and more generally in problems arising in random matrix theory \cite{Coquereaux2019} \cite{Coquereaux2019-2} \cite{Zuber2018} to  cite few recent results. We refer the reader to \cite{eynard:cea-01252029} for other results concerning this integral. One can consistently extend the definition of the HCIZ integral to all $\beta >0$ via the theory of rational Dunkl operator associated to generic root system \cite{anker:hal-01402334}, also known as the  \emph{Bessel hypergeometric function} in such setting. We list in the following interesting properties of the HCIZ integral: 

\paragraph{}
From its definition (\ref{HCIZdef}), one may notice that the HCIZ integrals only depends on the eigenvalues of its entries. In the following, we denote by $\bm{a}= \left(a_1, \dots, a_N \right)$  the vector of eigenvalues of the matrix $\mathbf{A}$, and $\underline{\bm{a}} = \Diag \left(a_1, \dots, a_N \right)$ the corresponding diagonal matrix, in particular the identity matrix is denoted by $\underline{\bm{1}}$.  For a given vector of eigenvalues $\bm{a}$, we see the HCIZ integral as a function of the vector $\bm{z}$, and we denote it by $\mathcal{I}^{(\beta)}_{\bm{a}}  \left( \bm{z} \right) $. It is also clear from its definition that the integral is invariant under the permutation of the vectors $\bm{a}$ and $\bm{z}$  and does not depend on the order of the entries of  each vector. It also satisfies: 

\begin{itemize}
\item For two diagonal matrices $\underline{\bm{a}}$ and $\underline{\bm{b}}$  and a matrix $\mathbf{G}$, if we denote by   $\bm{c}_{(\mathbf{G})} := \Eigen\left( \underline{\bm{a}} +  \mathbf{G} \underline{\bm{b}} \mathbf{G}^* \right) $, we have by Haar property:
    \begin{align}
\label{PropHCIZ}
\int \mathcal{D}\mathbf{G} \,  \mathcal{I}^{(\beta)}_{\bm{c}_{(\mathbf{G})} } \left( \bm{z} \right)  = \mathcal{I}^{(\beta)}_{\bm{a}} \left(\bm{z}\right) \, \mathcal{I}^{(\beta)}_{\bm{b}} \left(\bm{z} \right) &&
\end{align}
\item The HCIZ integral is normalized so that: \begin{align} \label{HCIZnorm}
 \mathcal{I}^{(\beta)}_{\bm{0} } \left( \bm{z} \right) &= 1 
\end{align}
\end{itemize}
For $\beta =2$, the HCIZ integral admits a nice determinantal formula due to Ityzkson and Zuber \cite{Itzykson1980}: 
 \begin{align} \label{HCIZ_det_form}
 \mathcal{I}^{(2)}_{\bm{a} } \left( \bm{z} \right) &= \left( \prod_{j=1}^{N-1} j! \right) \frac{\det \left[ e^{a_j z_k} \right]_{j,k=1}^N}{\Vand \left(\bm{a} \right) \Vand \left( \bm{z} \right) }
\end{align}where $\Vand(.) $ denotes the \emph{Vandermonde product}.

 The link between the HCIZ integral and the $R$-transform of the previous paragraph is given by the following property, first derived by Parisi \cite{Marinari1994} and then proved rigorously by Guionnet and Ma\"{\i}da  \cite{guionnet-maida} (see \cite{potters-bouchaud} for an introduction): 

\paragraph{}
If we denote by $\mathcal{I}^{(\beta)}_{\bm{a}}(z) := \mathcal{I}^{(\beta)}_{\bm{a}} \left(  z, 0, \dots, 0 \right) $ the rank-one specialization of the HCIZ integral, and if the spectral distribution satisfies the conditions of the previous paragraph, then for $z$ small enough, we have:

\begin{align}
\label{PGM}
\lim_{N \to \infty}\frac{2}{N \beta } \frac{d}{dz} \ln \mathcal{I}^{(\beta)}_{\bm{a}} \left( \frac{N \beta}{2} z\right) = \mathcal{R}_{\mu_A}(z) && 
\end{align}
\paragraph{}
Note that Parisi and Guionnet and Ma\"{\i}da actually prove that this result can be extended for finite rank $k$ and arbitrary large $z$ in the following sense: if we denote by $H^R_{\mu_A}(.)$ the unique function satisfying : 
\begin{align} \label{def_Hr}
\frac{d}{dz} H^R_{\mu_A}(z) =\begin{cases}
                \begin{array}{llll}
                 \mathcal{R}_{\mu_A}(z) && \text{ for }  0 \leq z \leq \mathcal{G}_{\mu_A} \left( a_{\max} \right) \\
                   a_{\max} - \frac{1}{z}&& \text{ for }  z \geq\mathcal{G}_{\mu_A} \left( a_{\max} \right)\\
        \end{array}
              \end{cases}
\end{align}and with $H^R_{\mu_A}(0)=0$. where $a_{\max}$ denotes the right edge of the limiting measure. Then we have: 
\begin{align}
\lim_{N \to \infty}\frac{2}{N \beta } \ln \mathcal{I}^{(\beta)}_{\bm{a}} \left( \frac{N \beta}{2} z_1, \dots, \frac{N \beta}{2} z_k , 0 , \dots, 0 \right) = \sum_{i=1}^k H^R_{\mu_A}(z_i)   && 
\end{align}
\paragraph{}
The goal of this note is to derive the multiplicative analogue of the theorem (\ref{PGM}), which to the best knowledge of the authors, as not been yet established and is therefore not well known. In Section 2, we introduce the spherical functions on symmetric cones, which are the multiplicative counterparts of the HCIZ integral in the case $\beta = 1,2,4 $ and then derived the asymptotic behavior of such functions by a saddle point analysis in this setting. In Section 3, we extend the derivation to all $\beta >0$ using recent results concerning Macdonald polynomials and Heckman-Opdam hypergeometric functions.

\section{Group integrals representations for $\beta =1, 2, 4 $ and asymptotic behavior }

\subsection{The spherical integral over symmetric cones}
\label{SecGrpInt}

Unlike the HCIZ integral, its multiplicative counterpart has been far less studied in the random matrix community, exception made of \cite{Zhang19}  \cite{Kieburg2019-2}  \cite{KK16} and references therein. We recall in this paragraph the construction of such function.

In the following the vector $\bm{a}$ (and $\bm{b}$) are assumed to be positive (that is for all $i$ we have $a_i >0$). The idea is to replace in  (\ref{PropHCIZ}) and (\ref{HCIZnorm}) the additive operation and its null element by the multiplicative operation and its null element. Namely we look a function $\mathcal{J}^{(\beta)}_{\bm{a}} \left( \bm{z} \right)$ such that 

\begin{itemize}
\item  for $\bm{c}_{(\mathbf{G})} = \Eigen\left( \underline{\bm{a}} \mathbf{G} \underline{\bm{b}} \mathbf{G}^* \right)$ \footnote{the eigenvalues of this product are the same as the ones of the free symmetric product}

\begin{align}
\label{propMul}
\int \mathcal{D} \mathbf{G} \, \mathcal{J}^{(\beta)}_{\bm{c}_{(\mathbf{G})}} \left( \bm{z} \right) =  \mathcal{J}^{(\beta)}_{\bm{a}} \left( \bm{z} \right) \, \mathcal{J}^{(\beta)}_{\bm{b}} \left( \bm{z} \right)
\end{align}
\item and normalized by: \begin{align} \label{normHO}
\mathcal{J}^{(\beta)}_{\bm{1}} \left( \bm{z} \right) = 1
\end{align}
\end{itemize}
\paragraph{}
It is well known from group theory, that if we take $\beta =2$ and we specialize the vector $\bm{z}$ to be a \emph{partition}, that is a vector of non-increasing non-negative integer, denoted by $\bm{\lambda}$, then the function $\mathcal{J}^{(2)}_{\bm{a}} \left( \bm{\lambda} \right)$ can be chosen to be  the classical \emph{Schur polynomials} \cite{macdonald2015}  $\mathrm{s}_{\bm{\lambda}} \left(\bm{a} \right) $  normalized by $s_{\bm{\lambda}}( 1,\dots, 1 )$.  Following \cite{macdonald2015}, the normalized Schur polynomials admits the following representation: 

\begin{align}
\label{normschur}
\frac{\mathrm{s}_{\bm{\lambda}} \left(\bm{a} \right)}{\mathrm{s}_{\bm{\lambda}} \left(1 , \dots,1 \right)} &= \int \mathcal{D}\mathbf{U}   \, \Delta_{\bm{\lambda}}(\mathbf{U} \underline{\bm{a}} \mathbf{U}^*)
\end{align}
where  for a vector $\bm{x}$ (which is not necessarily a partition) $\Delta_{\bm{x}}(.)$ is the \emph{multivariate power function} given by: 

\begin{align}
\label{defMPF}
\Delta_{\bm{x}} \left( \mathbf{A} \right) := \left( \det \mathbf{A}_{(1)} \right)^{x_1 - x_2} \dots  \left( \det \mathbf{A}_{(N-1)} \right)^{x_{N-1} - x_N } \left(   \det \mathbf{A} \right)^{ x_N }
\end{align}   where $\mathbf{A}_{(j)}$ is the $j \times j$ (top left) principal corner of the matrix $\mathbf{A}$. 
\paragraph{}
Now for $\beta =1 $ (resp. $\beta = 4$), with $\bm{\lambda}$ a partition, $\mathcal{J}^{(\beta)}_{\bm{a}} \left( \bm{\lambda} \right)$   can be chosen to be the normalized  real (resp. quaternionic) \emph{zonal polynomial} \cite{macdonald2015} which satisfies (\ref{normschur}) when one replaces the integral over the unitary group by an integral over the orthogonal (resp. symplectic) group. Since the multivariate power function $\Delta_{\bm{x}}(.)$ can defined for a general complex vector $\bm{x}$, this leads us to the following natural definition for the so-called \emph{multiplicative spherical function}: 

\begin{align}
\label{def_HO_grp}
\mathcal{J}^{(\beta)}_{\bm{a}} \left( \bm{z} \right) &:= \int \mathcal{D}\mathbf{G}  \,\Delta_{\bm{z}} (\mathbf{G}\underline{\bm{a}} \mathbf{G}^* ) 
\end{align}
with $\Delta_{\bm{x}}(.)$ defined in (\ref{defMPF}).  

\paragraph{Remark: } Using relationships between determinants of blocks of a matrix and of blocks of its inverse, one can show that $\mathcal{J}^{(\beta)}_{\bm{a}}(.)$ has the following symmetry: 

\begin{align} \label{prop_sym_inv}
\mathcal{J}^{(\beta)}_{\frac{1}{\bm{a}}} \left( \bm{z} \right) &:= \mathcal{J} _{\bm{a}}^{(\beta)}\left( - \sigma(\bm{z}) \right) 
\end{align}
where $\frac{1}{\bm{x}} := (\frac{1}{x_1}, \dots, \frac{1}{x_N})$ and where $\sigma(.)$ reverses the order of a vector (that is the permutation which exchange the $i^{th}$ argument with the $(N+1-i)^{th}$ argument). When $\bm{z}$ is a partition, this property allows us to define the corresponding symmetric (Laurent) polynomials for a \emph{signature}, that is a vector of non-decreasing integers (not necessarily positive).

\paragraph{}
Starting from (\ref{def_HO_grp}), it is possible to show that $\mathcal{J}^{(\beta)}_{\bm{a}} \left( \bm{z} \right)$ satisfies (\ref{propMul}) by use of the $QR$-decom\-position and the properties of the multivariate power function \cite{jacquesfaraut1995}. Unlike the HCIZ integral, this spherical function does not admit a double matrix integral representation.  Note also, to have similar property concerning symmetry and Dunkl/Calogero-Moser Operator, it is customary to  look at a \emph{shifted} version in $\bm{z}$ of this function, this is the so-called \emph{Heckman-Opdam hypergeometric function} of Section \ref{HOsec}.

\subsection{ Asymptotic behavior of the rank one spherical integral for $\beta = 1, 2, 4$  }
\label{As_beta124}

As in the additive case, we introduce the rank one specification as  $\mathcal{J}_{\bm{a}} \left( z \right) := \mathcal{J}_{\bm{a}} \left( z , 0 , \dots, 0 \right) $\footnote{unlike the HCIZ integral, this multiplicative function is not invariant by permutation in its argument $\bm{z}$ and the position of the non-trivial component of the rank one case matters. To have a permutation invariant function, one needs to  introduce a shift, giving the function of Section \ref{HOsec}.}, which is given by 
\begin{align}
\mathcal{J}^{(\beta)}_{\bm{a}}  \left( z \right) &= \int \mathcal{D}\mathbf{G} \, \det \left[ \left(\mathbf{G} \underline{\bm{a}} \mathbf{G}^* \right)_{(1)} \right]^{z}  && \left(\text{for } \beta = 1,2,4\right) 
\end{align}
By first projecting along the first row $\bm{x}$  of the group matrix $\mathbf{G}$ and then using the usual inverse Laplace representation of the constraint $\bm{x}^*\bm{x} = 1 $, we get:

\begin{align}
\mathcal{J}^{(\beta)}_{\bm{a}}  \left( z \right) &\propto\int_{\mathbb{F}^N} d\bm{x} \frac{1}{2 \pi \mathbf{i}} \int_{\gamma' - \mathbf{i} \infty}^{\gamma' + \mathbf{i} \infty} dt \, \, e^{t \left(1-\bm{x}^* \bm{x} \right)} \left( \bm{x}^* \underline{\bm{a}} \bm{x}\right)^z &&
\end{align}
where $\gamma'$ is a constant such that the complex integral along the vertical line is convergent and may change from line to line and $\mathbb{F} = \mathbb{R} , \mathbb{C} , \mathbb{H}$ for $\beta = 1, 2 , 4$ respectively. The constant of proportionality can be deduced using (\ref{normHO}) and is equal to $\frac{\Gamma \left( \frac{N\beta  }{2} \right)}{\pi^{\frac{N\beta}{2}}} $. 
Using the inverse Laplace representation of the power function for $a>0$:

\begin{align}
a^{z} = \frac{\Gamma(z+1)}{2 \pi \mathbf{i}} \int_{\gamma- \mathbf{i} \infty}^{\gamma+ \mathbf{i}  \infty }ds \, s^{-z-1} e^{sa}
\end{align}
we arrive at:

\begin{align}
\mathcal{J}^{(\beta)}_{\bm{a}}  \left( z \right) &=  \frac{\Gamma(z+1) \Gamma \left( \frac{N \beta  }{2} \right) }{ \pi^{\frac{N\beta  }{2}} } \frac{1}{2 \pi \mathbf{i}}  \int_{\gamma' - \mathbf{i} \infty}^{\gamma' + \mathbf{i} \infty} dt \, e^{t} \int_{\mathbb{F}^N} d\bm{x}   \, \, e^{-t \bm{x}^* \bm{x} }  \frac{1}{2 \pi \mathbf{i}} \int_{\gamma- \mathbf{i} \infty}^{\gamma+ \mathbf{i}  \infty }ds \, s^{-z-1} e^{s\left( \bm{x}^* \underline{\bm{a}} \bm{x} \right)}
\end{align}
by the change of variable $s = e^{-p}$  and deformation of the Bromwich contour, we get:

\begin{align}
\mathcal{J}^{(\beta)}_{\bm{a}}  \left( z \right) &= \frac{\Gamma(z+1) \Gamma \left( \frac{N\beta}{2} \right) }{ \pi^{\frac{N \beta  }{2}} }  \frac{1}{2 \pi \mathbf{i}} \int_{\gamma' - \mathbf{i} \infty}^{\gamma' + \mathbf{i} \infty} dt \, e^{t} \int_{\mathbb{F}^N} d\bm{x}  \, e^{-t \bm{x}^* \bm{x}} \frac{1}{2 \pi \mathbf{i}} \int_{\gamma- \mathbf{i} \infty}^{\gamma+ \mathbf{i}  \infty }dp \, e^{pz} e^{e^{-p}\left( \bm{x}^* \underline{\bm{a}} \bm{x} \right)} &&\\
\mathcal{J}^{(\beta)}_{\bm{a}}  \left( z \right) &= \frac{\Gamma(z+1) \Gamma \left( \frac{N \beta }{2} \right) }{ \pi^{\frac{N \beta }{2}} }  \frac{1}{2 \pi \mathbf{i}} \int_{\gamma' - \mathbf{i} \infty}^{\gamma' + \mathbf{i} \infty} dt \, e^{t}  \int_{\gamma- \mathbf{i} \infty}^{\gamma+ \mathbf{i}  \infty }dp  \left( \int_{\mathbb{F}^N} d\bm{x}  \, e^{- \bm{x}^* \left( t\underline{\bm{1}} - e^{-p} \underline{\bm{a}}  \right) \bm{x} } \right)  \, e^{pz} 
\end{align}

the term in bracket is a Gaussian integral so that after the change of variable $p \to p - \ln t$ and factorizing by $t^{-N}$ 

\begin{align}
\mathcal{J}^{(\beta)}_{\bm{a}}  \left( z \right) &=   \frac{\Gamma \left(z+1\right) \, \Gamma \left( \frac{N \beta }{2} \right)}{2 \pi \mathbf{i}} \int_{\gamma' - \mathbf{i} \infty}^{\gamma' + \mathbf{i} \infty} dt \, e^{t} t^{-\frac{\beta N }{2} -z}  \frac{1}{2 \pi \mathbf{i}}\int_{\gamma- \mathbf{i} \infty}^{\gamma+ \mathbf{i}  \infty }dp  \prod_{i=1}^N (1 - e^{-p} a_i )^{-\frac{\beta}{2}}\, e^{pz} 
\end{align}

and finally we arrive at:

\begin{align}
\mathcal{J}^{(\beta)}_{\bm{a}}  \left( z \right) &= \frac{\Gamma(1+z) \Gamma\left( \frac{N \beta}{2} \right)}{\Gamma(\frac{N \beta}{2}+z)} \mathcal{L}^{-1} \left[ \prod_{i=1}^N \left( 1 - e^{-p} a_i \right)^{-\frac{\beta}{2}} \right]\left( z\right)
\end{align}

To establish the Guionnet-Ma\"{\i}da theorem, we make the usual rescaling $z \to \frac{\beta N }{2} z$ and take the log derivative: 
\begin{align}
\frac{2}{N \beta}\frac{d}{dz} \ln \mathcal{J}^{(\beta)}_{\bm{a}} \left(\frac{N \beta}{2} z \right) =& \frac{2}{\beta} \frac{1}{N} \frac{d}{dz} \ln \left( \frac{\Gamma\left( \frac{N\beta}{2} \right) \Gamma \left(1 + \frac{N\beta}{2}z \right)}{\Gamma \left(\frac{N\beta}{2}+ \frac{N \beta}{2}z \right)} \right)
\notag\\
&
+ \frac{2}{N \beta}\frac{d}{dz} \ln\mathcal{L}^{-1} \left[\prod_{i=1}^N \left( 1- a_i e^{-p}  \right)^{- \frac{\beta}{2}} \right]
\end{align}
Taking the $N$ limit and by property of the digamma function $\psi(z) := \frac{d}{dz} \ln \Gamma(z) $ at infinity $\psi(z) \sim_{z \to \infty} \ln z  $, we have for the first term: 

\begin{align}
\lim_{N \to \infty} \frac{2}{N\beta}  \frac{d}{dz} \ln \left( \frac{\Gamma\left( \frac{N\beta}{2} \right) \Gamma \left(1 + \frac{N\beta}{2}z \right)}{\Gamma \left(\frac{N\beta}{2}+ \frac{N \beta}{2}z \right)} \right) = \ln{\frac{z}{z+1}}
\end{align}
while the second term is obtained by a saddle point method since: 

\begin{align}
\mathcal{L}^{-1} \left[\prod_{i=1}^N \left( 1- a_i e^{-p}  \right)^{- \frac{\beta}{2}} \right]\left(\frac{N \beta}{2} z \right) = \int dp \, e^{\frac{\beta N}{2} \mathcal{H}(z,p) }
\end{align}

with: 
\begin{align}\label{RateFunc}
\mathcal{H}(z,p) :=& zp - \int \mu_{\bm{a}}(dx) \ln \left( 1 - x e^{-p}\right)
\end{align}
from which we deduce that integral is dominated by  the critical point $p^*$ solution of: 
\begin{align}\label{eq_CriticalpointbeforeInversion}
z = \int \mu_{\bm{a}}(dx) \frac{x}{e^{p^*}-x}
=:\mathcal{T}_{\mu_A}(e^{p^*})
\end{align}
where the second equality is the definition of the $T$-transform. We then have:
\begin{align}
p^* = \ln \mathcal{T}_{\mu_A}^{(-1)}(z)
\end{align}
Since $p^*$ is an critical point, by total derivative of (\ref{RateFunc}) with respect to $z$, we conclude using (\ref{ModS-tsfrm}): 

\begin{align}\label{rank1S}
\Aboxed{
 \lim_{N \to \infty} \frac{2}{N \beta}\frac{d}{dz} \ln \mathcal{J}^{(\beta)}_{\bm{a}}  \left( \frac{N \beta}{2} z \right)  = \ln \tilde{\mathcal{S}}_{\mu_A}(z)  
 }  && \left(\text{for } \beta = 1,2,4\right) 
\end{align}
\paragraph{Remark:} We would like to point out that a similar limit in  the $\beta=2$  case has been carried out in \cite{gorin2018gaussian} but the authors did not write (\ref{rank1S}) explicitly.\\

The above argument is only valid for positive $z$ small enough so that the limiting $T$-transform is invertible. For larger $z$, one need to work with the discrete $T$-transform, in this case we find that $p^* = \ln a_{\max}$ where $a_{\max}$ is the largest element of $\bm{a}$. If we denote by $H^S_{\mu_A} (.)$ the solution of: 

\begin{align} \label{def_Hs}
\frac{d}{dz} H^S_{\mu_A}(z) =\begin{cases}
                \begin{array}{llll}
               \ln\mathcal{S}_{\mu_A}(z) && \text{ for }  0 \leq z \leq   \mathcal{T}_{\mu_A} \left( a_{\max} \right) \\
               \ln a_{\max} + \ln \left(   \frac{z}{z+1} \right) && \text{ for }  z \geq \mathcal{T}_{\mu_A} \left(  a_{\max} \right)\\
        \end{array}
              \end{cases}
 \end{align}
 with $H^S_{\mu_A}(0) =0$, then we have: 

\begin{align}\label{allz}
 \lim_{N \to \infty} \frac{2}{N \beta} \ln \mathcal{J}^{(\beta)}_{\bm{a}}  \left( \frac{N \beta}{2} z , 0 , \dots, \, 0 \right)  =  H^S_{\mu_A} (z_i)
 \end{align}Note that from (\ref{def_Hs}) and the power series expansion of the $S$-transform \cite{potters-bouchaud}, one can compute explicitly  the first coefficients of the power series of $H^S_{\mu_A}(.)$ near the origin,  in terms of the first moments of the distribution $\mu_A$,which we denote by $m_k$. One has: 
 
 \begin{align}
  H^S_{\mu_A}(z) = \left( \ln m_1 \right) z + \left( \frac{m_2 }{m_1^2} -1\right) \frac{z^2}{2} + \left(  \frac{  2 m_3 m_1 -3 m_2^2 }{m_1^4}  +1 \right)\frac{z^3}{6} + O(z^4)
 \end{align}

\paragraph{Remark:}
Using non-rigorous arguments that we expect to be valid in the large $N$ limit, we convinced ourselves that this result should
also hold in the finite $k$ rank setting then one should have:
\begin{align}\label{low_rank_conj}
 \lim_{N \to \infty} \frac{2}{N \beta} \ln \mathcal{J}^{(\beta)}_{\bm{a}}  \left( \frac{N \beta}{2} z_1 , \dots, \frac{N \beta}{2} z_k  , 0 , \dots, 0 \right)  = \sum_{i=1}^k H^S_{\mu_A} (z_i)
\end{align}

Note that the since spherical integral \eqref{def_HO_grp} is not permutation invariant on the elements of $\bm{z}$, it matters which elements of $\bm{z}$ are non-zero in our low-rank limit. As long as the non-zero elements are all below some finite position $n$ as $N$ goes to infinity, we expect our permutation invariant conjecture \eqref{low_rank_conj} to hold true.

\section{Extension to all $\beta >0$: Heckman-Opdam hypergeometric function}

\subsection{Heckman-Opdam and the spherical integral}
\label{HOsec}

 The Harish-Chandra integral can be extended to all $\beta >0$, by abstracting out its group integral representation thanks to the theory of rational Dunkl operators associated to root system of type $A_{N-1}$\cite{anker:hal-01402334} \cite{Amri2014}. A similar construction can be performed in the multiplicative case for the so-called \emph{Heckman-Opdam hypergeometric function} \cite{CM_1987__64_3_329_0} \cite{opdam2000} by use of the trigonometric Dunkl theory \cite{anker:hal-01402334}, such function is the permutation-invariant symmetric version (in its argument $\bm{z}$) of the spherical integral of the previous section.  We first give its group integral representation in the $\beta =1 ,2,4$ and show how one can naturally extend it to all $\beta >0$, without involving the theory of root systems. 
\paragraph{}
For $\beta = 1, 2, 4$ the Heckman-Opdam is a shifted version of the spherical function of Section \ref{SecGrpInt}, namely: 

\begin{align}
\label{def_HO_sph}
\mathcal{F}^{(\beta)}_{\bm{a}} \left( \bm{z} \right) &:= \mathcal{J}^{(\beta)}_{e^{-\bm{a}}} \left( -\bm{z} - \bm{\rho} \right) && \left(\text{for } \beta = 1,2,4\right) 
\end{align}
with $e^{\bm{x}} = \left( e^{x_1}, \dots, e^{x_N} \right) $ and $\bm{\rho}$  given by: 

\begin{align}
\rho_i& := \frac{\beta}{2} \left( N - i \right) \quad \text{for } i = 1,\dots, N
\end{align}This operation is equivalent to put an additional $\frac{\beta}{2}$ term in the exponent of each determinant in (\ref{defMPF}), except for the last one. 

\paragraph{Remark:} Using the symmetry relation (\ref{prop_sym_inv}) and the permutation invariance of the Heckman-Opdam hypergeometric, this can also be written as:
\begin{align} \label{def_Ho_2}
\mathcal{F}^{(\beta)}_{\bm{a}} \left( \bm{z} \right) &:= \mathcal{J}^{(\beta)}_{e^{\bm{a}}} \left( \bm{z} + \sigma(\bm{\rho}) \right) && \left(\text{for } \beta = 1,2,4\right) 
\end{align}
with $ \left[\sigma(\bm{\rho}) \right]_i = \frac{\beta}{2}(i-1)$. 
\paragraph{}

In particular for $\beta =2$, the Heckman-Opdam hypergeometric function admits  a multiplicative counterpart of the Itzykson-Zuber determinantal formula (\ref{HCIZ_det_form}), known as the  Gelfand-Naimark \cite{GelfandNaimark} formula: 

 \begin{align}
 \mathcal{F}^{(2)}_{\bm{a}} \left( \bm{z} \right) = \left( \prod_{j=1}^{N-1} j! \right) \frac{\det \left[  e^{a_j z_k}\right]^N_{j,k=1} }{\Vand \left( e^{-\bm{a}}\right) \Vand  \left( -\bm{z}\right) } = \left( \prod_{j=1}^{N-1} j! \right) \frac{ \det\left( e^{\bm{a}}\right)^{N-1}   \det \left[  e^{a_j z_k}\right]^N_{j,k=1} }{\Vand \left( e^{\bm{a}}\right) \Vand  \left( \bm{z}\right) }
 \end{align}where $\Vand( .)$ is the usual Vandermonde determinant. 
 \paragraph{}
The HCIZ integral can be obtain as a limit of the Heckman-Opdam hypergeometric function \cite{Gorin2018}  \cite{BenSad2005} namely we have: 
\begin{align}
\mathcal{I}^{(\beta)}_{\bm{a} } \left( \bm{z} \right) = \lim_{\epsilon \to 0^+ } \mathcal{F}^{(\beta)}_{ \epsilon \bm{a}} \left( \epsilon^{-1}\bm{z} \right)
\end{align}
\subsection{The spherical integral for arbitrary beta}
 The spherical integral and the Heckman-Opdam function are defined using the Haar measure on the orthogonal/unitary/simplectic group. We will show here how to generalize their definition to arbitrary beta. The goal is to write the joint distribution of eigenvalues of all principal minors of randomly rotated fixed matrix. Consider first  the principal minor $ \mathbf{M}$ of size $N-1$ of a rotated matrix $\mathbf{A}$ of size $N$, its eigenvalues are equivalent to the non-zero eigenvalues of 
 \begin{equation}
     \mathbf{M}=\mathbf{\Pi}\mathbf{A}\mathbf{\Pi}\qquad \mathbf{\Pi}=\underline{\bm{1}}-\bm{x} \bm{x}^*
 \end{equation}
where $\underline{\bm{1}}$ is the identity matrix and $\bm{x}$ is a normalized Gaussian vector whose statistics can  easily be generalized to arbitrary $\beta$. The non-zero eigenvalues $\{\lambda_i\}$ of $\mathbf{M}$ satisfy the interlacing condition $ a_i\geq \lambda_{i} \geq a_{i+1}$ where we have assumed that the eigenvalues of $\mathbf{A}$ are in decreasing order.
The joint law of $\{\lambda_i\}$ is given by the Dixon-Anderson integral \cite{Dixon1905} \cite{Neretin2003} (see chapter 4 of \cite{Forrester} for a derivation)
\begin{equation}\label{dixon-anderson}
    P_{\bm{a}}^\beta(\{\lambda_i\})=
    \frac{\Gamma(\frac{N\beta}2)}{\Gamma(\frac{\beta}2)^N}
    \left(\prod_{1\leq i\leq j \leq N-1}|\lambda_i-\lambda_j|\right)\left(\prod_{1\leq i\leq j \leq N}|a_i-a_j|^{1-\beta}\right)\prod_{i=1}^{N-1}\prod_{j=1}^N |\lambda_i-a_j|^{\frac{\beta}2-1}
\end{equation}
one can then iterate the procedure to obtain the joint law of eigenvalues of all principal minors, known as the \emph{the beta corner process}  \cite{Gorin2018}. When applied  to (\ref{def_HO_sph}), 
where we have reordered each vector $\bm{a}$ such that it is decreasing,
we obtain the following expression for $\mathcal{F}^{(\beta)}_{\bm{a}} \left( \bm{z} \right) $:  
\begin{align}
\mathcal{F}^{(\beta)}_{\bm{a}} \left( \bm{z} \right) =& \frac{1}{Z_{N,\beta,e^{-\bm{a}}}}\int \prod_{j=1}^N e^{a_j z_N} \prod_{k=1}^{N-1} \prod_{i=1}^k \left(\lambda^{(k)}_i \right)^{-(z_k - z_{k+1})} \prod_{k=1}^{N-1} \prod_{i=1}^k \left(\lambda^{(k)}_i \right)^{-\frac{\beta}{2}} 
\notag\\
&\times \prod_{k=1}^{N-1} \left(\prod_{1 \leq i <j \leq k} |\lambda^{(k)}_i - \lambda^{(k)}_j |^{2-\beta} \right) \left( \prod_{u=1}^k \prod_{v=1}^{k+1} |\lambda^{(k)}_u -\lambda^{(k+1)}_v |^{\frac{\beta}{2}-1} \right) \prod_{k=1}^{N-1} \prod_{i=1}^k d\lambda^{(k)}_i 
\end{align}
where the integral is over the set of $\{ \lambda_i^{(k)} \}_{ 1 \leq i \leq k \leq N}$ satisfying the interlacing constraints $ \lambda_i^{(k+1)}\geq \lambda_{i}^{(k)} \geq \lambda_{i+1}^{(k+1)}$ and $\bm{\lambda}^{(N)} = e^{-\bm{a}}$, and

\begin{align}
\label{Normcst_BCP}
Z_{N,\beta,e^{-\bm{a}}} := (-1)^{\frac{N(N+1)}{2}}\prod_{k=1}^N \frac{\Gamma(\frac{\beta}{2})^k}{\Gamma\left( \frac{k \beta}{2} \right)} \prod_{1 \leq i < j \leq N } ( e^{-a_j} - e^{-a_i} )^{\beta -1} 
\end{align}

If we make the change of variable $\lambda^{(k)}_i = e^{-l^{(k)}_i}$, with $\{l_i^{(k)}\}$ satisfying the interlacing constraints with $\bm{l}^{(N)}=\bm{a}$, this introduces a constant $(-1)^{\frac{N(N+1)}{2}}$which exactly cancel out with the one in  (\ref{Normcst_BCP}). We have:
\begin{align}
\label{Def_HO_GT}
\mathcal{F}^{(\beta)}_{\bm{a}} \left( \bm{z} \right) =& \frac{\prod_{k=1}^N \Gamma(\frac{k\beta}{2})}{ \Gamma(\frac{\beta}{2})^{\frac{N(N+1)}{2}} \prod_{1 \leq i < j \leq N } ( e^{-a_j} - e^{-a_i} )^{\beta -1} }\int e^{\sum_{k=1}^N z_k \left( \sum_{i=1}^k l^{(k)}_i - \sum_{i=1}^{k-1} l^{(k-1)}_i \right)} 
\notag\\
&\times \prod_{k=1}^{N-1} \left(\prod_{1 \leq i <j \leq k} | e^{-l^{(k)}_i} -  e^{-l^{(k)}_j} |^{2-\beta} \right) \left( \prod_{u=1}^k \prod_{v=1}^{k+1} | e^{-l^{(k)}_u} - e^{-l^{(k+1)}_v} |^{\frac{\beta}{2}-1} \right)  \prod_{i=1}^k   e^{\left(\frac{\beta}{2} - 1\right) l^{(k)}_i} dl^{(k)}_i 
\end{align}
From this formula it is  possible to establish a link between the Heckman-Opdam hypergeometric function and the normalized \emph{Macdonald polynomials} and we refer the reader to \cite{Sun16} \cite{Sun2016} \cite{Borodin2014} for a derivation of the result. The \emph{Macdonald polynomials} $  \mathrm{P}_{\bm{\lambda}} \left(. | q ,t\right) $ are $q,t$-deformation of the Schur polynomials and we refer to \cite{macdonald2015} for an introduction on this subject.  We have:
\begin{align}
\label{prop_HO_McD}
\mathcal{F}^{(\beta)}_{ \bm{a}}(\bm{z}) = \lim_{\epsilon \to 0^+} \frac{\mathrm{P}_{\lfloor \frac{ \bm{a}}{\epsilon} \rfloor} \left(e^{\epsilon \bm{z}}  | e^{- \epsilon} , e^{- \epsilon\frac{\beta}{2}}\right)}{\mathrm{P}_{\lfloor \frac{ \bm{a}}{\epsilon} \rfloor}\left(1, e^{- \epsilon\frac{\beta}{2}} ,\dots, e^{- \epsilon\frac{\beta}{2} \left(N-1\right)}|e^{- \epsilon} , e^{- \epsilon\frac{\beta}{2}} \right) }
\end{align}
where $e^{\bm{x}} = \left( e^{x_1} , \dots, e^{x_N} \right)$ and   $\lfloor \bm{x} \rfloor := \left( \lfloor x_1 \rfloor , \dots, \lfloor x_N \rfloor \right) $ with $\lfloor . \rfloor$  the integer part function.  Our interest for this expression lies in the fact that the normalized Macdonald polynomials admits a simple integral representation when all its arguments except one are fixed. 

\subsection{Rank one formula and asymptotic behavior}

We recall that we want to extend the asymptotic behavior of the rank one function $\mathcal{J}^{(\beta)}_{\bm{a}} \left( z,0, \dots,0 \right)$ of Section \ref{SecGrpInt} to all $\beta >0$, which is equivalent to the study of $\mathcal{F}^{(\beta)}_{- \ln  \bm{a}}(-z - (N-1)\frac{\beta}{2}, \dots, -\frac{\beta}{2} )$ by (\ref{def_HO_sph}). The negative-value partition, also known as a \emph{signature}, Macdonald polynomials are defined by (\cite{Gorin2018}): 

\begin{align}
\mathrm{P}_{-\bm{\lambda}} \left(\bm{x} |q,t \right) = \mathrm{P}_{\bm{\lambda}} \left(\bm{x}^{-1} |q,t\right) 
\end{align}
and by homogeneity of the Macdonald polynomials and (\ref{prop_HO_McD}) we are left with the study of \footnote{Note that we have assume the vector $\bm{a}$ to be non decreasing so that the index of the Macdonald polynomial in (\ref{prop_HO_McD}) is a well-defined partition} \footnote{this is clear also clear using (\ref{def_Ho_2})}: 

\begin{align}
\label{rk1_HO_Mcd}
\mathcal{J}^{(\beta)}_{\bm{a}}  \left( z \right) :=  \lim_{\epsilon \to 0^+}  \frac{\mathrm{P}_{\lfloor \frac{\ln \bm{a}}{\epsilon} \rfloor} \left(e^{\epsilon z},  e^{- \epsilon\frac{\beta}{2}} ,\dots, e^{- \epsilon\frac{\beta}{2} \left(N-1\right)}  | e^{- \epsilon} , e^{- \epsilon\frac{\beta}{2}}\right)}{\mathrm{P}_{\lfloor \frac{ \ln \bm{a}}{\epsilon} \rfloor}\left(1, e^{- \epsilon\frac{\beta}{2}} ,\dots, e^{- \epsilon\frac{\beta}{2} \left(N-1\right)}|e^{- \epsilon} , e^{- \epsilon\frac{\beta}{2}} \right) } && \left(\text{for } \beta > 0 \right) 
\end{align}

It turns out that the corresponding specification of the normalized Macdonald polynomials appearing in (\ref{rk1_HO_Mcd}) admits a simple formula in this setting \cite{Cuenca2018}. For a signature $\bm{\lambda}$, $q \in (0,1) $ and  a complex $|x| >1$, we have:

\begin{align}
\label{Rk1Mcd}
\frac{\mathrm{P}_{\bm{\lambda}} \left(x, 1,  q^{\frac{\beta}{2}} ,\dots, q^{\frac{\beta}{2} \left(N-2\right)}  | q,  q^{\frac{\beta}{2}}\right)}{\mathrm{P}_{\bm{\lambda}}\left(1,q^{\frac{\beta}{2}} ,\dots, q^{\frac{\beta}{2} \left(N-1\right)}| q,  q^{\frac{\beta}{2}}\right) } = & \frac{\ln q}{q-1}\Gamma_{ q} \left(\frac{\beta N}{2} \right) \frac{\left(\frac{ q^{\frac{N\beta}{2}}}{x} ; q \right)_{\infty} }{\left( \frac{q}{x} ; q\right)_{\infty}}
\times \frac{1}{2 \pi \mathbf{i}} \int_{C}dp \, x^{p}  \prod_{i=1}^N \frac{\Gamma_{q} \left( p - \left( \lambda_i - \frac{\beta}{2}  i + \frac{\beta}{2}  \right) \right)}{\Gamma_{ q} \left( p - \left( \lambda_i - \frac{\beta}{2} i \right) \right)}  
\end{align}

where $\Gamma_q \left(z\right):=  (1-q)^{1-z}\frac{\left( q ; q \right)_{\infty}}{\left( q^z  ;q \right)_{\infty}} $ is the $q$-gamma function and $\left(. ; q \right)_{\infty}$ the $q$-Pochhammer of $q$-calculus \cite{Andrews1999} and $C$ is a usual complex Bromwich contour which left all the poles to the left of the integration line. Without loss of generality we fix $\mathfrak{Re} z >0$, the other case can be obtain using the symmetry relation (\ref{prop_sym_inv}) and the behavior of the $S$-transform of the inverse of a matrix \cite{potters-bouchaud}.  By doing the change of variables $p \to \epsilon p$ , we have:

\begin{align}
\mathcal{J}^{(\beta)}_{\bm{a}}  \left( z \right)  &= \lim_{\epsilon \to 0^+} \frac{- \epsilon}{1 - e^{-\epsilon}} \Gamma_q \left(\frac{\beta N}{2} \right)  \frac{\left(e^{-\epsilon (\frac{N\beta}{2}+ z)} ; e^{-\epsilon} \right)_{\infty} }{\left(e^{-\epsilon (1+  z)} ; e^{-\epsilon} \right)_{\infty}} \left( 1 - e^{- \epsilon} \right)^{\frac{N\beta}{2}} \frac{1}{\epsilon}
\notag\\
& \times \frac{1}{ 2 \pi \mathbf{i}} \int_{C}dp \, e^{p z} \prod_{i=1}^{N}\frac{\left(e^{\epsilon \lfloor \frac{\ln a_i}{\epsilon} \rfloor} e^{-p} \, e^{- \epsilon \frac{\beta}{2}i } ; e^{- \epsilon} \right)_{\infty}}{\left( e^{\epsilon \lfloor \frac{\ln a_i}{\epsilon} \rfloor}  e^{-p} \, e^{- \epsilon \frac{\beta}{2} (i-1) } ; e^{- \epsilon} \right)_{\infty}}  
\end{align}

Next taking the limit $\epsilon \to 0^+$ together with the following limit relation of $q$-calculus \cite{Andrews1999}: 
  
\begin{enumerate}
\item $\lim_{q \to 1^- } \Gamma_q \left( x \right) = \Gamma (x) $ 
\item $\lim_{q \to 1^- } \frac{\left(q^a ; q \right)_{\infty}}{\left(q^b ; q \right)_{\infty}} (1-q)^{a - b} = \frac{\Gamma(b)}{\Gamma(a)} $
\item $\lim_{q \to 1^- } \frac{\left(u \, q^a ; q \right)_{\infty}}{\left(u q^b ; q \right)_{\infty}} = \left( 1 - u \right)^{b-a} \qquad $ for $|u|<1$  
\end{enumerate}
and  $\lim_{\epsilon \to 0^+} e^{\epsilon \lfloor \frac{\ln a_i}{\epsilon}\rfloor}  = a_i  $, we have : 

\begin{align}
\mathcal{J}^{(\beta)}_{\bm{a}}  \left( z \right)  = \frac{\Gamma \left( \frac{\beta N }{2} \right) \Gamma \left( 1 + z  \right)}{\Gamma \left(\frac{N\beta}{2} + z \right) } \frac{1}{2 \pi \mathbf{i}} \int_{C^+}dp \,  e^{z p} \prod_{i=1}^N \left( 1- a_i e^{-p}  \right)^{- \frac{\beta}{2}} 
\end{align}

which is the generalization to all $\beta >0 $ of the previous formula: 

\begin{align}
\mathcal{J}^{(\beta)}_{\bm{a}}  \left( z \right) = \frac{\Gamma \left( \frac{\beta N }{2} \right) \Gamma \left( 1 + z  \right)}{\Gamma \left(\frac{N\beta}{2} + z \right) } \mathcal{L}^{-1} \left[\prod_{i=1}^N \left( 1- a_i e^{-p}  \right)^{- \frac{\beta}{2}} \right](z)
\end{align}
and since the saddle-point analysis of Section \ref{As_beta124} still holds for $\beta >0$, we deduce that for z small enough (with positive real value), we have: 

\begin{align}\label{rk1S_allBETA}
\Aboxed{
 \lim_{N \to \infty} \frac{2}{N \beta}\frac{d}{dz} \ln \mathcal{J}^{(\beta)}_{\bm{a}}  \left( \frac{N \beta}{2} z \right)  = \ln \tilde{\mathcal{S}}(z)  
 }  && \left(\text{for } \beta > 0 \right) 
\end{align}

\section{Asymptotics of symmetric polynomials}
Just like the spherical function of Section \ref{SecGrpInt} can be seen as an extension in index of the Schur polynomials ($\beta =2$) and zonal polynomials ($\beta = 1 ,4 $), the Heckman-Opdam hypergeometric function can be seen as an extension of (a $\bm{\rho}$-shifted version of) the normalized \emph{Jack polynomials} $ \mathrm{j}_{\bm{\lambda}}^{\left(\frac{2}{\beta}\right)} \left( \bm{a} \right) $ \cite{Borodin2014} for all $\beta >0$ . For $\bm{\lambda}$ a partition, we have: 
\begin{align}
\mathcal{F}^{(\beta)}_{\bm{a}} \left(- \bm{\lambda} - \bm{\rho} \right) = \frac{\mathrm{j}_{\bm{\lambda}}^{\left(\frac{2}{\beta}\right)} \left( e^{-\bm{a}} \right) }{\mathrm{j}_{\bm{\lambda}}^{\left(\frac{2}{\beta}\right)} \left(1, \dots, 1 \right)}
\end{align}

For $\beta =2$, the asymptotic behavior of the HCIZ integral can be translated as an asymptotic behavior over normalized Schur polynomial: if $\frac{1}{N} \sum_{i=1}^N \delta_{ N^{-1}(\lambda_i + N -i)}$ converge toward a deterministic measure $\mu$, than the corresponding  normalized Schur polynomial with \emph{index} $\bm{\lambda}$ and with all  its arguments except one fixed, converges (up to an integration term) exponentially towards the integral of the $R$-transform (\ref{def_Hr}) see \cite{guionnet-maida}. Since the multiplicative spherical function $\mathcal{J}^{(\beta)}_{\bm{a}}  \left( z \right) $ of this note is nothing else than the analytical extension of the Jack polynomials $\mathrm{j}_{\bm{\lambda}}^{\left(\frac{2}{\beta}\right)} \left( .\right) $ we have a similar interpretation, except that now it is the vector in argument of the Jack polynomial which converges towards a deterministic measure while the index is the trivial partition $\bm{\lambda} = ( \lfloor \frac{N \beta}{2} z \rfloor , 0, \dots, 0 ) =: \lfloor \frac{N \beta}{2} z \rfloor $

\begin{align} \label{Asymp_jack}
\lim_{N \to \infty} \frac{2}{N\beta} \ln \frac{\mathrm{j}_{\lfloor \frac{N \beta}{2} z \rfloor  }^{\left(\frac{2}{\beta}\right)} \left(  \bm{a}\right)}{\mathrm{j}_{\lfloor \frac{N \beta}{2} z \rfloor  }^{\left(\frac{2}{\beta}\right)} \left( 1, \dots,1\right)} =  H^S_{\mu_A} (z)
\end{align}
with $H^S_{\mu_A} (.)$ defined by (\ref{def_Hs}). In particular, for $\beta =2$, the Jack polynomials become Schur polynomials and Schur polynomials of a trivial partition degenerate into \emph{complete homogeneous polynomials} defined by: 

\begin{align}
\mathrm{h}_{k} (\bm{a}) = \sum_{ 1\leq i_1 \leq \dots \leq i_k \leq N} a_{i_1} \dots a_{i_k}
\end{align}so that the LHS of (\ref{Asymp_jack}) has a simple explicit expression in terms of the $a_i$ in this case. 

\begin{figure}
  \centering
\includegraphics[scale=0.5]{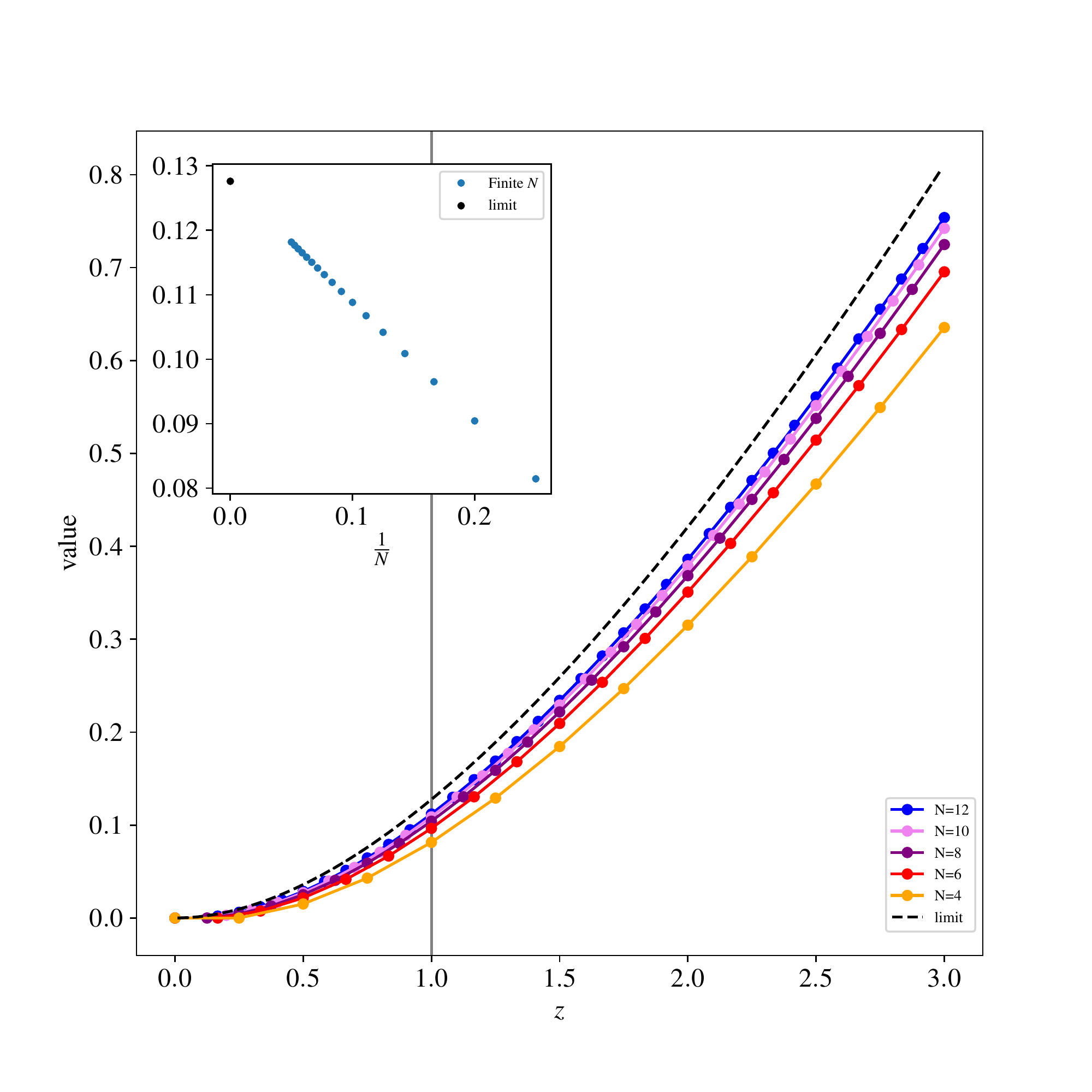}
\caption{ Value of the logarithm of the normalized complete homogeneous symmetric polynomials (\ref{Asymp_h})  for equidistributed entries between $0$ and $2$  for different $N$ and different $k=Nz$, compared with the limiting behavior (\ref{H_uni}), represented by a dashed line. The inset graph represents the convergence at $k=N$ ($z=1$) for more values of $N$, represented as a function of $\frac{1}{N}$. }
\label{fig_H_Com_SPol}
\end{figure}

\paragraph{}
As an illustration of this example, we take $\mu_{A}$ to be the uniform distribution between $0$ and $2$, then after some calculation one has: 

\begin{align} \label{H_uni}
H^S_{\mu_{A}}(z)  = z \left( \ln \frac{2z}{ \left| z+1 + W\left(-(z+1)e^{-(z+1)} \right) \right| } -1 \right) - \ln \left| W \left( -(z+1)e^{-(z+1)} \right) \right| 
\end{align}
where $W(.)$ is  the \emph{Lambert W function}, which we compare with:

\begin{align} \label{Asymp_h}
 \frac1N\ln \frac{\mathrm{h}_{k}\left(  \bm{a}\right)}{\mathrm{h}_{k} \left( 1, \dots,1\right)}
 =\frac1N\ln\left[ \frac{k!(N-1)!}{(k+N-1)!}\mathrm{h}_{k}\left(  \bm{a}\right)\right]
\end{align}

 for different $N$ and $z=k/N$, where the $a_i$  are the $N$ equidistributed points between $0$ and $2$. The results are shown in Fig. \ref{fig_H_Com_SPol}.
 
 Note that the conjecture (\ref{low_rank_conj}) implies that Jack polynomials with few non-trivial entries in the index partition should completely decoupled in this regime. In particular ``low-rank'' Schur polynomials should converge to a product of complete homogeneous symmetric polynomials.

\section*{Conclusion}
In this note we have studied the asymptotic behavior of the multiplicative spherical function and its link with the $S$-transform, establishing the multiplicative counterpart of the Parisi-Guionnet-Ma\"\i da theorem. Our result is expected to be true in the large $N$ limit when the fraction of non-zero entries in the vector $\mathbf{z}$  goes to zero. For the HCIZ integral there exists an asymptotic regime where both matrices are full rank and their eigenvalues converge to well determined measures \cite{matytsin1994large}. It would be quite interesting to find out if the Heckman-Opdam function (or equivalently the spherical integral) can be computed asymptotically in the regime where both vectors $\mathbf{z}$ and $\mathbf{a}$ converge to full measures.

\newpage
\bibliographystyle{unsrt}

\bibliography{biblio}

\begin{thebibliography}{10}

\bibitem{HarishChandra1957}
Harish-Chandra.
\newblock Differential operators on a semisimple {Lie} algebra.
\newblock {\em American Journal of Mathematics}, 79(1):87, 1957.

\bibitem{Itzykson1980}
Claude Itzykson and Jean-Bernard Zuber.
\newblock The planar approximation. {II}.
\newblock {\em Journal of Mathematical Physics}, 21(3):411--421, 1980.

\bibitem{McSwiggen2018}
Colin McSwiggen.
\newblock A new proof of {Harish}-{Chandra}'s integral formula.
\newblock {\em Communications in Mathematical Physics}, 365(1):239--253, 2019.

\bibitem{Goulden2014}
Ian~P. Goulden, Mathieu Guay-Paquet, and Jonathan Novak.
\newblock Monotone {Hurwitz} numbers and the {HCIZ} integral.
\newblock {\em Annales math{\'{e}}matiques Blaise Pascal}, 21(1):71--89, 2014.

\bibitem{novak2020complex}
Jonathan Novak.
\newblock On the complex asymptotics of the {HCIZ} and {BGW} integrals, 2020.

\bibitem{Coquereaux2019}
Robert Coquereaux and Jean-Bernard Zuber.
\newblock {The Horn Problem for Real Symmetric and Quaternionic Self-Dual
  Matrices}.
\newblock {\em {Symmetry, Integrability and Geometry : Methods and
  Applications}}, 15:029, 2019.

\bibitem{Coquereaux2019-2}
Robert Coquereaux, Colin McSwiggen, and Jean-Bernard Zuber.
\newblock Revisiting {Horn}'s problem.
\newblock {\em Journal of Statistical Mechanics: Theory and Experiment},
  2019(9):094018, 2019.

\bibitem{Zuber2018}
Jean-Bernard Zuber.
\newblock {Horn}'s problem and {Harish}-{Chandra}'s integrals. probability
  density functions.
\newblock {\em Annales de l'Institut Henri Poincaré}, 5(3):309--338, 2018.

\bibitem{eynard:cea-01252029}
Bertrand Eynard, Taro Kimura, and Sylvain Ribault.
\newblock {\em Random matrices}.
\newblock arXiv:1510.04430, 2015.

\bibitem{anker:hal-01402334}
Jean-Philippe Anker.
\newblock {An introduction to Dunkl theory and its analytic aspects}.
\newblock In G.~Filipuk, Y.~Haraoka, and S.~Michalik, editors, {\em {Analytic,
  Algebraic and Geometric Aspects of Differential Equations}}, Trends in
  Mathematics, pages 3--58. {Birkh{\"a}user}, 2017.

\bibitem{Marinari1994}
Enzo Marinari, Giorgio Parisi, and Felix Ritort.
\newblock Replica field theory for deterministic models. {II}. a non-random
  spin glass with glassy behaviour.
\newblock {\em Journal of Physics A: Mathematical and General},
  27(23):7647--7668, 1994.

\bibitem{guionnet-maida}
Alice Guionnet and Myl{\`e}ne Ma{\"\i}da.
\newblock A {Fourier} view on the {R}-transform and related asymptotics of
  spherical integrals.
\newblock {\em Journal of Functional Analysis}, 222(2):435 -- 490, 2005.

\bibitem{potters-bouchaud}
Marc Potters and Jean-Philippe Bouchaud.
\newblock {\em A first course in random matrix theory}.
\newblock Cambridge University Press, (in press) 2021.

\bibitem{Zhang19}
Jiyuan Zhang, Mario Kieburg, and Peter Forrester.
\newblock Harmonic analysis for rank-1 randomised {Horn} problems.
\newblock arXiv:1911.11316, 2019.

\bibitem{Kieburg2019-2}
Mario Kieburg, Peter~J. Forrester, and Jesper~R. Ipsen.
\newblock Multiplicative convolution of real asymmetric and real anti-symmetric
  matrices.
\newblock {\em Advances in Pure and Applied Mathematics}, 10(4):467--492, 2019.

\bibitem{KK16}
Mario Kieburg and Holger Kösters.
\newblock Exact relation between singular value and eigenvalue statistics.
\newblock {\em Random Matrices: Theory and Applications}, 05(04):1650015, 2016.

\bibitem{macdonald2015}
Ian~G. Macdonald.
\newblock {\em Symmetric Functions and {Hall} Polynomials (Oxford Classic Texts
  in the Physical Sciences: Oxford Mathematical Mongraphs)}.
\newblock Oxford University Press, 2015.

\bibitem{jacquesfaraut1995}
Jacques Faraut.
\newblock {\em Analysis on Symmetric Cones (Oxford Mathematical Monographs)}.
\newblock Clarendon Press, 1995.

\bibitem{gorin2018gaussian}
Vadim Gorin and Yi~Sun.
\newblock Gaussian fluctuations for products of random matrices, 2018.

\bibitem{Amri2014}
B{\'{e}}chir Amri.
\newblock Note on {Bessel} functions of type {$A_{N-1}$}.
\newblock {\em Integral Transforms and Special Functions}, 25(6):448--461,
  2014.

\bibitem{CM_1987__64_3_329_0}
Gert~J. Heckman and Eric~M. Opdam.
\newblock Root systems and hypergeometric functions. {I}.
\newblock {\em Compositio Mathematica}, 64(3):329--352, 1987.

\bibitem{opdam2000}
Eric~M. Opdam.
\newblock {\em Part I: Lectures on Dunkl Operators}, volume Volume 8 of {\em
  MSJ Memoirs}, pages 2--62.
\newblock The Mathematical Society of Japan, Tokyo, Japan, 2000.

\bibitem{GelfandNaimark}
Israel~M. Gelfand and Mark~A. Na\u{i}mark.
\newblock Unitary representations of the classical groups.
\newblock {\em Trudy Mat. Inst. Steklov.}, 36:288, 1950.

\bibitem{Gorin2018}
Vadim Gorin and Adam~W Marcus.
\newblock Crystallization of random matrix orbits.
\newblock {\em International Mathematics Research Notices}, 2018.

\bibitem{BenSad2005}
Salem~Ben Saïd and Bent {\O}rsted.
\newblock Analysis on flat symmetric spaces.
\newblock {\em Journal de Math{\'{e}}matiques Pures et Appliqu{\'{e}}es},
  84(10):1393--1426, 2005.

\bibitem{Dixon1905}
Arthur~L. Dixon.
\newblock Generalization of {Legendre}'s formula {$KE' -(K-E)K' = \frac{1}{2}
  \pi $}.
\newblock {\em Proceedings of the London Mathematical Society},
  s2-3(1):206--224, 1905.

\bibitem{Neretin2003}
Yu~A Neretin.
\newblock Rayleigh triangles and non-matrix interpolation of matrix beta
  integrals.
\newblock {\em Sbornik: Mathematics}, 194(4):515--540, 2003.

\bibitem{Forrester}
Peter~J. Forrester.
\newblock {\em Log gases and Random Matrices}.
\newblock Princeton University Press, 2010.

\bibitem{Sun16}
Yi~Sun.
\newblock Matrix models for multilevel {H}eckman-{O}pdam and multivariate
  {B}essel measures.
\newblock arXiv:1609.09096, 2016.

\bibitem{Sun2016}
Yi~Sun.
\newblock A new integral formula for {Heckman}{\textendash}{Opdam}
  hypergeometric functions.
\newblock {\em Advances in Mathematics}, 289:1157--1204, 2016.

\bibitem{Borodin2014}
Alexei Borodin and Vadim Gorin.
\newblock General $\beta$-{Jacobi} corners process and the gaussian free field.
\newblock {\em Communications on Pure and Applied Mathematics},
  68(10):1774--1844, 2014.

\bibitem{Cuenca2018}
Cesar Cuenca.
\newblock Asymptotic formulas for {M}acdonald polynomials and the boundary of
  the $(q, t)$-{G}elfand-{T}setlin graph.
\newblock {\em Symmetry, Integrability and Geometry: Methods and Applications},
  14:66, 2017.

\bibitem{Andrews1999}
George~E. Andrews, Richard Askey, and Ranjan Roy.
\newblock {\em Special Functions}.
\newblock Cambridge University Press, 1999.

\bibitem{matytsin1994large}
A~Matytsin.
\newblock On the large-{N} limit of the {Itzykson}-{Zuber} integral.
\newblock {\em Nuclear Physics B}, 411:805--820, 1994.

\end{thebibliography}
\end{document}